\documentclass[twocolumn,aps,prb,showpacs]{revtex4}
\usepackage{graphicx}

\begin{document}

%
%

\title{Exchange effect and magneto-plasmon mode dispersion
in an anisotropic two-dimensional electronic system}

\author{Xiaoguang Wu}

\affiliation{SKLSM, Institute of Semiconductors, Chinese Academy
of Sciences, Beijing 100083, China}

\begin{abstract}

The exchange effect and the magneto-plasmon mode dispersion are studied
theoretically for an anisotropic two-dimensional electronic system in
the presence of an uniform perpendicular magnetic field.  Employing an
effective low-energy model with anisotropic effective masses, which
is relevant for a monolayer of phosphorus, the exchange effect due to
the electron-electron interaction is treated within the self-consistent
Hartree-Fock approximation.  The magneto-plasmon mode dispersion is
obtained by solving a Bethe-Salpeter equation for the electron
density-density correlation function within the ladder diagram approximation.
It is found that the exchange effect is reduced in the anisotropic system
in comparison with the isotropic one.  The magneto-plasmon mode dispersion
shows a clear dependence on the direction of the wave vector.

\end{abstract}


\maketitle

%
%

\section{Introduction}

In recent years, the black phosphorus has attracted great attention
because of its interesting physical properties and its great potential in the
electronic and electro-optical device applications.\cite{rev01,rev02}

Many experimental works are devoted to the material growth,
the physical property characterization, and the device
exploration.\cite{rev01,rev02,exp01,exp02,exp03,exp04}
There is also a considerable amount of theoretical investigations
that concerning the electronic band structures,\cite{t01,t02,t03}
the Landau levels and the anisotropic optical properties,\cite{t04,t05,t06}
the plasmon at zero magnetic field,\cite{t07,t08,t09,t10}
the topological and edge states,\cite{t11,t12}
the anisotropic composite fermions,\cite{t13,t14},
the electron substrate phonon coupling,\cite{t15}
and the tuning of the band gap by a bias electric field.\cite{t16}
It is interesting to note that the monolayer black phosphorus
may provide an alternative two-dimensional electronic system to
study the influence of interplay between the anisotropy and
the electron-electron interaction.  This interplay has been studied
in a GaAs quantum well recently.\cite{gaas}  In these
theoretical investigations, we feel that the many-body effects
arising from the electron-electron interaction is less extensively
studied.  More studies on the many-particle effect is desired.

This motivates the present study.  In this paper, we will study
the exchange effect in an anisotropic two-dimensional electronic
system, in the presence of a perpendicular magnetic field, due
to the electron-electron interaction within the self-consistent
Hartree-Fock approximation.  This exchange effect is an important
ingredient for the explanation of the quantum oscillation displayed
in the magneto-transport property of an isotropic two-dimensional
electronic system.  For an anisotropic system, one expects that
it is important as well.  We will also study the magneto-plasmon
mode dispersion beyond the random-phase-approximation.
The magneto-plasmon was observed experimentally for the
isotropic two-dimensional electronic system.\cite{batke}

The present paper is organized as follows: in the section II,
the approach used is briefly presented.  In the section III,
our theoretical results are shown and discussed.
Finally, a brief summary is provided in the last section.

\section{Formulations and calculations}

In this paper, the anisotropic two-dimensional (2D) electronic
system in the $xy$ plane is modelled by the following Hamiltonian
$$ H = {{1}\over{2m_1}} ( p_x + {{eA_x}\over{c}} )^2
     + {{1}\over{2m_2}} ( p_y + {{eA_y}\over{c}} )^2
     + g^* \mu_B B s_z
\phantom{...}, $$
in the presence of an uniform perpendicular magnetic field $B$.
This effective low-energy Hamiltonian can be used to describe the
conduction band of a monolayer of phosphorus.\cite{t01,t05}
The valence band is omitted for simplicity and clarity.  The 2D
electronic system displays an anisotropy via two different effective
masses $m_1$ and $m_2$.  $g^*$ is the effective $g$-factor,
and $s_z$ is the electron spin operator.
The electron-electron interaction is assumed to be isotropic and
the Coulomb potential is a function of the distance between two electrons.

By choosing a gauge as ${\bf A}=(0,Bx,0)$, one obtains
the single-particle energy level ($\sigma=\pm 1$)
$$ \varepsilon^{0}_{l,\sigma} = \hbar \omega_c (l + 1/2)
   + g^* \mu_B B \sigma/2
\phantom{...}, $$
with $\omega_c = eB/(c \sqrt{m_1 m_2})$.  The Landau levels are
degenerated, and this degeneracy leads to an area
density $n_B=eB/(2\pi\hbar c)=1/(2\pi l_B^2)$, independent of the
effective mass parameters.  The corresponding one-particle wave
function is given by
$$ \phi_{l,k_y}({\bf r}) = [e^{ik_y y}/\sqrt{L_y}]
   [ \sqrt{\alpha}
   \psi_{l}(\alpha(x-x_0)) ]
\phantom{...}, $$
with $x_0 = l_B^2 k_y$, $\alpha=(m_1 \omega_c/\hbar)^{1/2}$, and
$$ \psi_{n}(x) = [\pi^{1/2} 2^n n!]^{-1/2} e^{-x^2/2} H_n(x)
\phantom{...}, $$
with $H_n(x)$ the Hermit polynomial.\cite{merzbacher}
Note that the length scale $\alpha$ depends on the effective
mass parameter.

It is interesting to note that the well-known
Kohn's theorem is also valid for this anisotropic 2D electronic
system.\cite{kohn}  This leads to a constrain on the magneto-plasmon
mode dispersion at small wave vectors.

The exchange effect due to the electron-electron interaction is
treated within the self-consistent Hartree-Fock approximation.\cite{fetter,mahan}
The imaginary time Green's function can be written as
$$ G^0_{\sigma}({\bf r}, {\bf r}', i\omega_n)
   = \sum_{l,k_y}
   {{\phi_{l,k_y}({\bf r}) \phi^*_{l,k_y}({\bf r}')}
   \over{i\omega_n - (\varepsilon^0_{l,\sigma} - \mu)/\hbar}}
\phantom{...}, $$
for the non-interacting system.\cite{fetter}  After taking into
account the exchange effect, the Green's function becomes
$$ G_{\sigma}({\bf r}, {\bf r}', i\omega_n)
   = \sum_{\lambda,k_y}
   {{\psi_{\lambda,k_y}({\bf r}) \psi^*_{\lambda,k_y}({\bf r}')}
   \over{i\omega_n - (\varepsilon_{\lambda,\sigma} - \mu)/\hbar}}
\phantom{...}, $$
with $\psi_{\lambda,\sigma}({\bf r})=\sum_m C_{\lambda,\sigma; m}
\phi_{m,k_y}({\bf r})$.  The Landau level energy $\varepsilon_{\lambda,\sigma}$,
with the exchange effect included, is still independent of $k_y$.
The expansion coefficients $C_{\lambda,\sigma; m}$ needs to be
calculated self-consistently.

The magneto-plasmon mode dispersion is obtained by solving a
Bethe-Salpeter equation for the electron density-density correlation
function within the ladder diagram approximation.\cite{fetter}
The electron density-density correlation function can be written as
$$ D({\bf q}, i\omega_n) = \sum_{\lambda}
   {{\pi_{\lambda}({\bf q})}\over{i\omega_n - \omega_{\lambda}({\bf q})}}
\phantom{...}, $$
where $\omega_{\lambda}({\bf q})$ gives the magneto-plasmon mode
dispersion, and $\pi_{\lambda}({\bf q})$ gives the corresponding oscillator strength.
The damping of magneto-plasmon can not be obtained in this approach.
For an isotropic system, the magneto-plasmon mode dispersion was
calculated previously.\cite{kallin,macdonald}  The effect of disorder
was studied.\cite{antoniou}  The influence of
multiple subbands and spin-orbit coupling was also considered
previously for the isotropic system.\cite{xgwu}

There is a simple scaling relation when the electron density-density
correlation function is calculated for an non-interaction 2D electronic
system.  By denoting the electron density-density correlation function
of an non-interacting isotropic 2D system, in the presence of a perpendicular
magnetic field, as $D^0_{\rm iso}(ql_B, \omega/\omega_c)$, then the
corresponding electron density-density correlation function for the anisotropic
case can be obtained as
$$ D^0_{\rm aniso}({\bf q}l_B, \omega/\omega_c)
   = D^0_{\rm iso}(p l_B, \omega/\omega_c)
\phantom{...}, $$
with $p_x=q_x(m_2/m_1)^{1/2}$ and $p_y=q_y(m_1/m_2)^{1/2}$.
This scaling relation can be easily verified by directly calculating
the one-bubble electron density-density correlation function.\cite{fetter}
With the help of this scaling relation, a qualitative picture can be
quickly drawn for the magneto-plasmon mode dispersion within the
random-phase-approximation from previously known theoretical result.

\section{Results and discussions}

In this paper, we will use $\hbar\omega_c$ as the energy scale, $l_B$
as the length scale.  The wave vector will be scaled by $1/l_B$.
The electron density enters our calculation via the filling
factor $\nu=n_e/n_B$ for a single spin.  As the Landau level energy
will be scaled by $\hbar\omega_c$, and the magneto-plasmon mode frequency
will be scaled by $\omega_c$, only the ratio of two effective
masses $m_1$ and $m_2$ is needed in our calculation.
However, the values of $m_1$ and $m_2$ used in the calculation for a
particular figure will be given individually without
causing any confusion.

For clarity, we will take the effective $g$-factor a vanishingly
small value.  However, the occupation of the Landau levels is assumed
to be spin resolved.  The spin-down Landau level is assumed to be
occupied first, and the spin-up Landau level to be occupied next.
This approximation scheme has to be modified when one considers the
case of a tilted and large magnetic field that the Zeeman spin
splitting may become larger than the separation of two neighboring
Landau levels.  The system temperature is assumed to be zero.

Another parameter enters our calculation is the strength of the
electron-electron interaction. This parameter is introduced via the
following dimensionless quantity $v_e=\sqrt{2} e^2/(\epsilon_s l_B)/(\hbar\omega_c)$,
with $\epsilon_s$ the background dielectric constant.  We takes
$v_e<1$ in accordance with the nature of the perturbation theory
employed in the present work.

In Fig.1, the energy of a few low Landau levels versus the filling
factor is plotted.  For an even filling factor of a value $2n$,
$n$ spin-down Landau levels and $n$ spin-up Landau levels are occupied.
This leads to the same exchange correction to the spin-down Landau
level and the spin-up Landau level.  In this case, the spin splitting
of the Landau level will be purely come from the Zeeman spin splitting.
As we choose $g^*=0$, there is no spin splitting of the Landau level
shown in the figure, as one expects.

For an odd filling factor of a value $2n+1$, there are $n+1$ spin-down Landau
levels occupied, and $n$ spin-up Landau levels occupied.  The $(n+1)$-th
spin-up Landau is empty.  This results in a larger energy shift, due to the
exchange effect, for the spin-down Landau level. Consequently, a spin-splitting
of the Landau level occurs.

When the filling factor takes a fractional
value, an integer number of Landau levels for one spin are fully occupied,
while for the other spin, one Landau level will be partially occupied.
Thus, one observes that the spin-down (spin-up) Landau level energy
remains constant, and the spin-up (spin-down) Landau level energy
decreases linearly as the filling factor increases from one integer
to the next higher integer.  This picture is qualitative the same as
in the case of an isotropic electronic system.\cite{kallin,macdonald,antoniou,xgwu}
In the Fig.1, the parameters used are $m_1=1$, $m_2=2.4$, and $v_e=0.61$.
Note that only the ratio $m_1/m_2$ matters as mentioned previously.
In a monolayer of phosphorus, the value of $m_1/m_2$ can be quite large,\cite{rev01}
we have chosen a moderate value instead.

In Fig.2, the energy of the lowest two Landau levels is shown versus
the effective mass ratio $m_2/m_1$.  There is an symmetry.  The energy
levels will be the same when $m_2/m_1=r$ and when $m_2/m_1=1/r$.
This reflects the fact that the physics must not
change when one denotes the direction with the effective mass $m_1$ as
the $x$-axis or as the $y$-axis.  One observes that the correction to
the Landau levels due to the exchange effect is largest when $m_2/m_1=1$.
This is because that when $m_2/m_1\ne 1$, the 2D electronic system has
a lower symmetry, the exchange effect leads to a mixing of Landau levels,
and this mixing effect reduces the energy shift due to the exchange interaction.
This can be viewed as the reduction of the exchange effect due to the
anisotropy.

One also observes that, the depth of two smile curves shown
in the Fig.2 is different.  This shows that the spin splitting of a
Landau level is also reduced due the anisotropy in a 2D electronic
system.  The above features shown for the lowest two Landau levels
also exist for the higher Landau levels.  In the Fig.2, the parameters
used are filling factor $\nu=3.1$ and $v_e=0.71$.

Next, let us examine the magneto-plasmon mode dispersion.
In Fig.3, the frequency of magneto-plasmon modes around $\omega_c$
versus the amplitude of the wave vector is plotted for three
different integer filling factors, $\nu=1,2,3$.
The other parameters used are $v_e=0.87$,
$m_1=1$, and $m_2=3$.  The direction of the wave vector is
characterized by an angle $\varphi$ ($q_x=q\cos(\varphi)$),
and it takes a fixed value
of $\varphi=\pi/6$ for the dispersion curves shown in the Fig.3.
The magneto-plasmon modes are determined by the poles in the
electron density-density correlation function.\cite{fetter,mahan}  There are many modes,
but in this paper, we will focus on the mode whose frequency is
around $\omega_c$.

In the case of $\nu=1$, only the $n=0$ spin-down Landau level is
occupied, the electron density excitation around $\omega_c$ should
arise from a transition between the $n=0$ spin-down Landau level
and the $n=1$ spin-down Landau level.  As the spin-up Landau levels
are empty, there is no spin-up Landau level transition.  Therefore,
one would expect only one magneto-plasmon mode around $\omega/\omega_c=1$.
In the cases of $\nu=2$ and $\nu=3$, following a similar argument,
one would expect two magneto-plasmon modes.  Our calculation
corroborates this hand waving picture.

Note that there is no spin-density
excitation shown, as the 2D electronic system studied here has
no spin-orbit interaction, and we are limited ourselves to the
electron density excitations.  The spin-density excitation can be
evaluated by summing up a series of slightly different ladder
diagrams.\cite{fetter,mahan}
When the filling factor is not an integer, the spin-density
excitation will show up as a collective excitation frequency
less than $\omega_c$, around the Zeeman spin splitting.

The magneto-plasmon modes can be classified into $E_1$ modes and
$E_2$ modes for small wave vectors.  They are indicated explicitly in the Fig.3.  For
small wave vectors, the amplitude of $E_1$ modes behaves like
$\pi_{\lambda}({\bf q})\sim q^2$.  The amplitude of $E_2$ modes is much
smaller.  In the long wave length limit, the shown $E_1$ modes dominate,
and the corresponding magneto-plasmon mode frequencies approach
$\omega_c$ as required by the Kohn's theorem.\cite{kohn}

For an isotropic 2D electronic system, the magneto-plasmon mode was
observed experimentally by depositing a grating coupler to a surface
close to the 2D electronic system.\cite{batke}  However, the wave
vector achieved at that time was small.  We wish that this theoretical
work will inspire more experimental investigations taking the advantage
of currently available more advanced chip technology.

The dispersions shown in the Fig.3 display a similar $q$-dependence
as that for an isotropic 2D electronic system.\cite{kallin}  However,
for an isotropic system, the magneto-plasmon mode frequency only
depends on the magnitude of the wave vector, and is independent of the
direction of the wave vector.  This is no longer the case for the
anisotropic 2D electronic system.  This will be examined next.

In Fig.4, the frequency of magneto-plasmon modes around $\omega_c$
is depicted versus the direction of the wave vector, characterized by $\varphi$ the
angle between the wave vector ${\bf q}$ and the $x$ axis, for the
integer filling factors $\nu=1,2,3$.  The magnitude of the wave vector
is fixed at $ql_B=0.75$.  The other parameters used are $v_e=0.87$, $m_1=1$,
and $m_2=3$.  One observes that the magneto-plasmon mode dispersion
shows a clear angular dependence.  As $\varphi$ changes, the
magneto-plasmon mode frequency displays an oscillatory behavior.
The amplitude of this oscillation depends on the magnitude of the
wave vector, but in a non-monotonic way.

Calculations are carried out for some $q$ values.  It is found
that $E_1$ modes have a larger $\varphi$ oscillations amplitude,
and the amplitude for the $E_2$ modes is smaller.  In the Fig.4,
one observes that, the $E_1$ magneto-plasmon mode frequency
takes its minimal value around $\varphi=\pi/2$ and $\varphi=3\pi/2$.
This behavior is only found for some small $q$ values.
For some larger $q$ values, the minimal is located around $\varphi=0$,
or falls between $\varphi=0$ and $\varphi=\pi/2$.

It should be pointed
out that, the Landau level energy will not change if one exchanges
the values of two effective masses $m_1$ and $m_2$.  However, the
magneto-plasmon mode dispersions shown in the Fig.3 and Fig.4, does
change if one swaps the values of $m_1$ and $m_2$.  This is expected,
as the wave function expansion in the Green's function is different
for different effective masses.

In Fig.5, we plot the frequency of magneto-plasmon modes versus
the amplitude of the wave vector, for a non-integer filling
factor $\nu=2.3$.  The other parameters used are $v_e=0.87$, $m_1=1$,
$m_2=3$, and $\varphi=\pi/6$.  For $\nu=2.3$, the occupation of the
Landau levels are as follows: $n=0$ spin-down and $n=0$ spin-up Landau
levels are fully occupied, $n=1$ spin-down Landau level is partially
occupied, and all other Landau levels are empty.  The electron density
excitations around $\omega_c$ could arise (1) from the transition
between the $n=0$ to $n=1$ spin-down Landau levels, (2) from the
transition between the $n=0$ to $n=1$ spin-up Landau levels, and
(3) from the transition between the $n=1$ to $n=2$ spin-down Landau
levels.  One expects $3$ modes.  In the Fig.5, one observes $E_1$
and $E_2$ modes as one observed in the Fig.3 for $\nu=2,3$.
A new mode appears and is labelled as $E_3$.  This mode is absent
in the Fig.3.  This mode arises from the third transition discussed
above.  We find that, at long wave length (small $q$), the dominate
mode is still $E_1$, the strength of $E_2$ and $E_3$ modes are much
smaller.  This is expected from the Kohn's theorem.\cite{kohn}

In Fig.6, the frequency of magneto-plasmon modes versus
the direction of the wave vector is plotted, for the non-integer
filling factor $\nu=2.3$, at $ql_B=0.75$.  Other parameters used are
the same as that in the Fig.5, $v_e=0.87$, $m_1=1$, and $m_2=3$.
One observes a clear angular dependence.  The magneto-plasmon mode
dispersion shows an oscillatory behavior versus $\varphi$.  The
amplitude of this oscillation is larger for the $E_1$ mode, and
smaller for the $E_2$ mode and $E_3$ mode.  We find that, this
behavior also holds for some other $q$ values.

The band structure of a monolayer of phosphorus is more complicated
than the simple anisotropic model used in the present study.\cite{rev01,rev02,t01,t16}
However, we believe that the picture shown here for the magneto-plasmon
mode will qualitatively remain when the influence of a more realistic
band structure is taken into account.  On the other hand, the influence
of disorder, of spin-orbit interaction, and of electron-phonon coupling
should be studied in the future.

%
%

\section{Summary}

In summary, we have used an effective low-energy model with two effective
masses to investigate the influence of anisotropy in an interacting
two-dimensional electronic system, relevant for a monolayer
of phosphorus.  The electron-electron interaction induced exchange effect
and the magneto-plasmon mode dispersion are studied theoretically.
The correction due to the exchange interaction to the Landau levels is
evaluated within the self-consistent Hartree-Fock approximation.
The magneto-plasmon mode dispersion is calculated by solving a Bethe-Salpeter
equation for the electron density-density correlation function within
the ladder diagram approximation, beyond the random phase approximation.
It is found that the exchange effect is
reduced in the anisotropic system in comparison with the isotropic one.
The magneto-plasmon mode dispersion shows a strong dependence on the
direction of the wave vector.

\acknowledgments{}

This work was partly supported NSF of China via
projects 61076092 and 61290303.

%
%

%
%

\begin{figure}[h]
\includegraphics[angle=90, width=8.truecm]{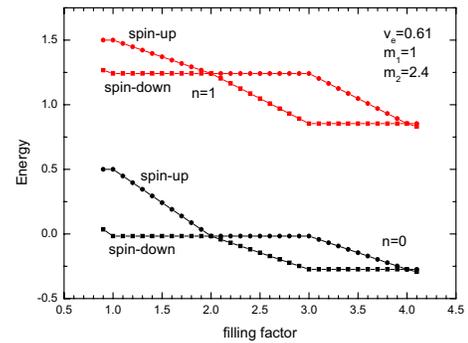}
\caption{ (Color online)
The energy of a few low-lying Landau levels versus the filling
factor. }
\end{figure}

\begin{figure}[h]
\includegraphics[angle=90, width=8.truecm]{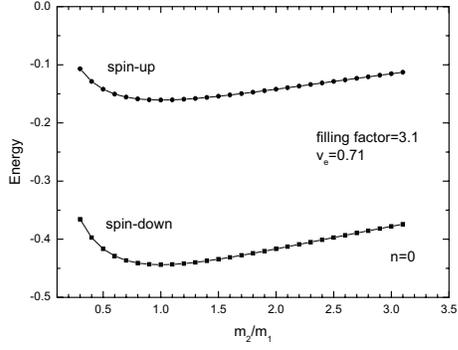}
\caption{ (Color online)
The energy of the lowest two Landau levels versus the effective mass
ratio. }
\end{figure}

\begin{figure}[h]
\includegraphics[angle=90, width=8.truecm]{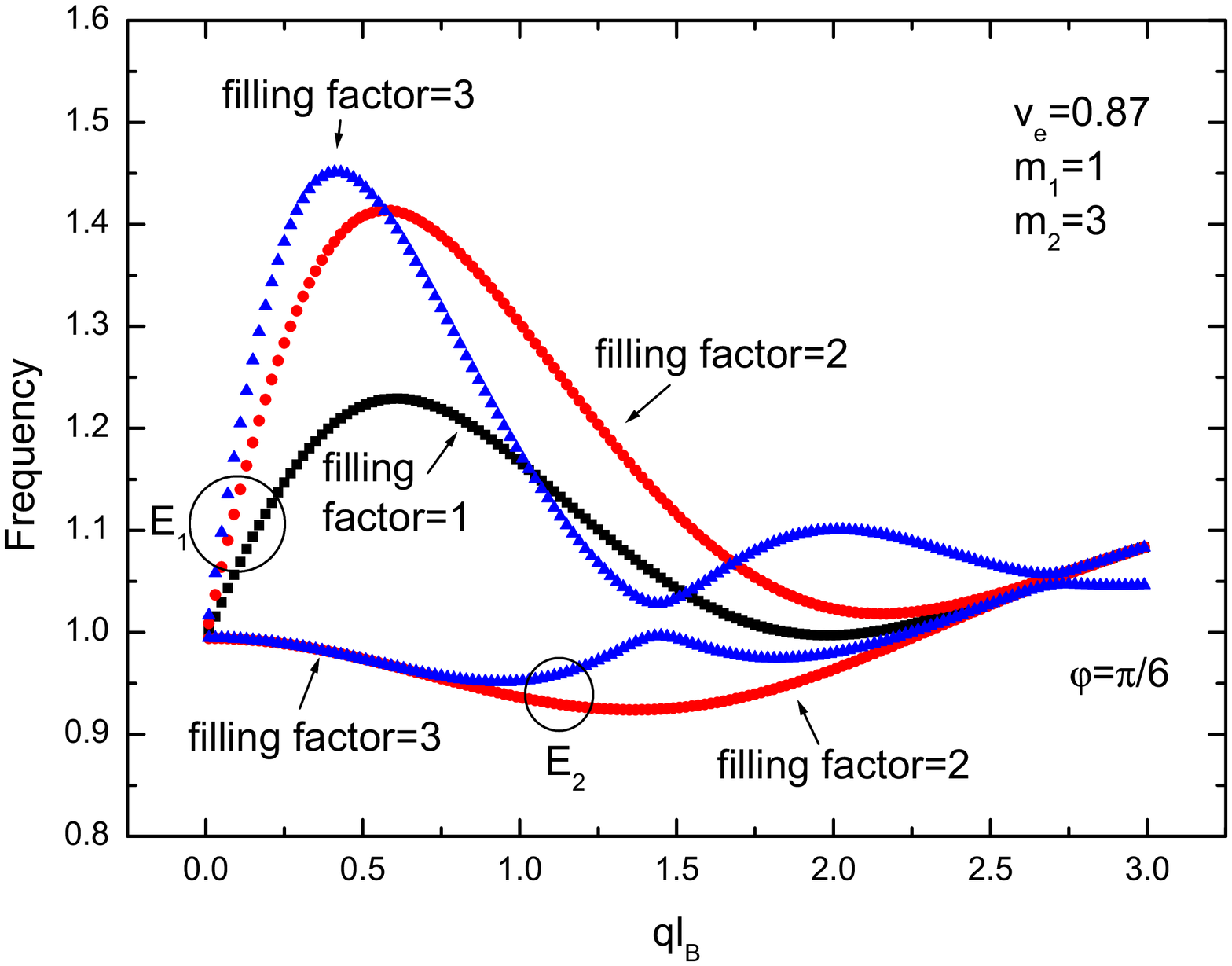}
\caption{ (Color online)
The frequency of magneto-plasmon modes versus the amplitude
of the wave vector, for some integer filling factors. }
\end{figure}

\begin{figure}[h]
\includegraphics[angle=90, width=8.truecm]{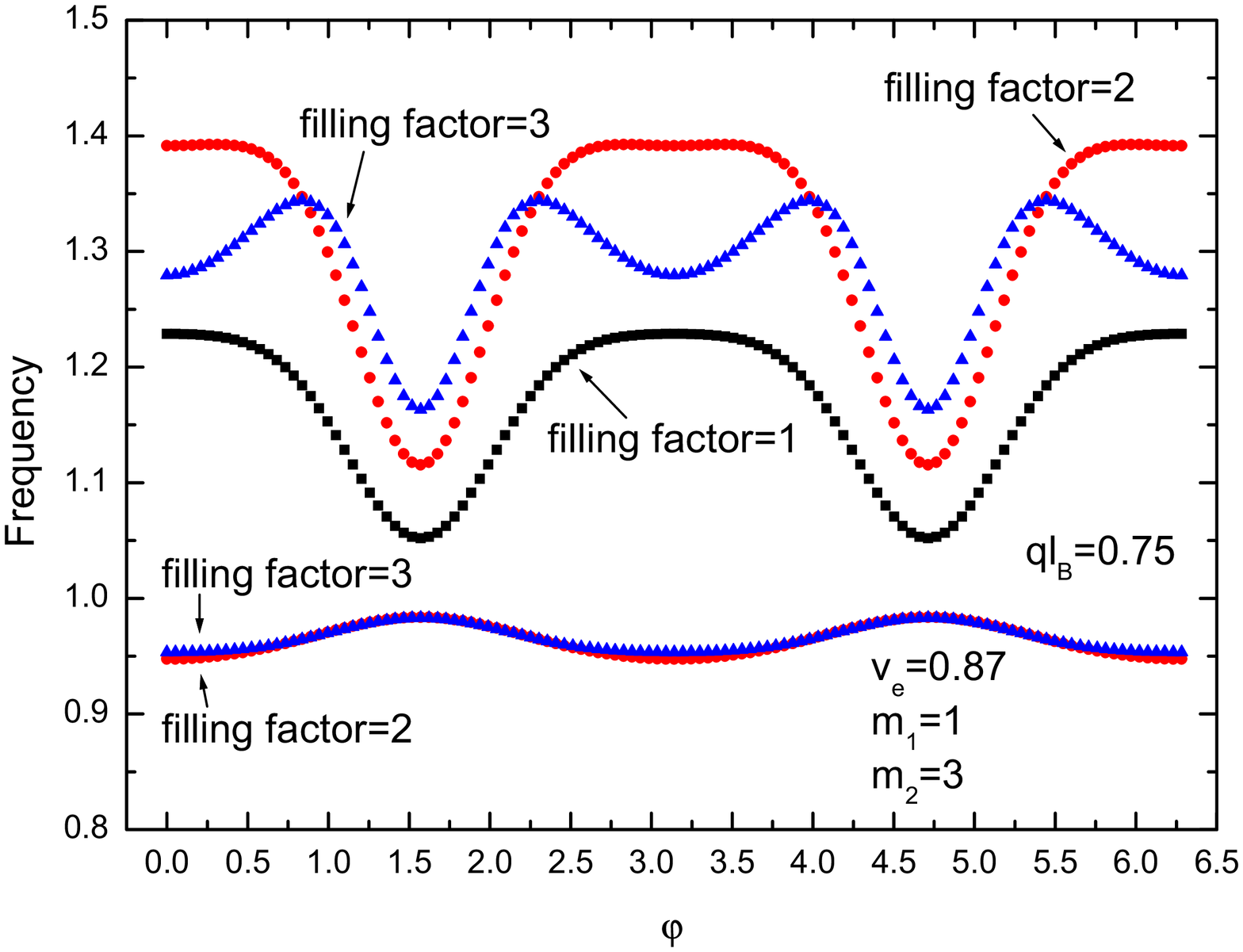}
\caption{ (Color online)
The frequency of magneto-plasmon modes versus the direction
of the wave vector, for some integer filling factors. }
\end{figure}

\begin{figure}[h]
\includegraphics[angle=90, width=8.truecm]{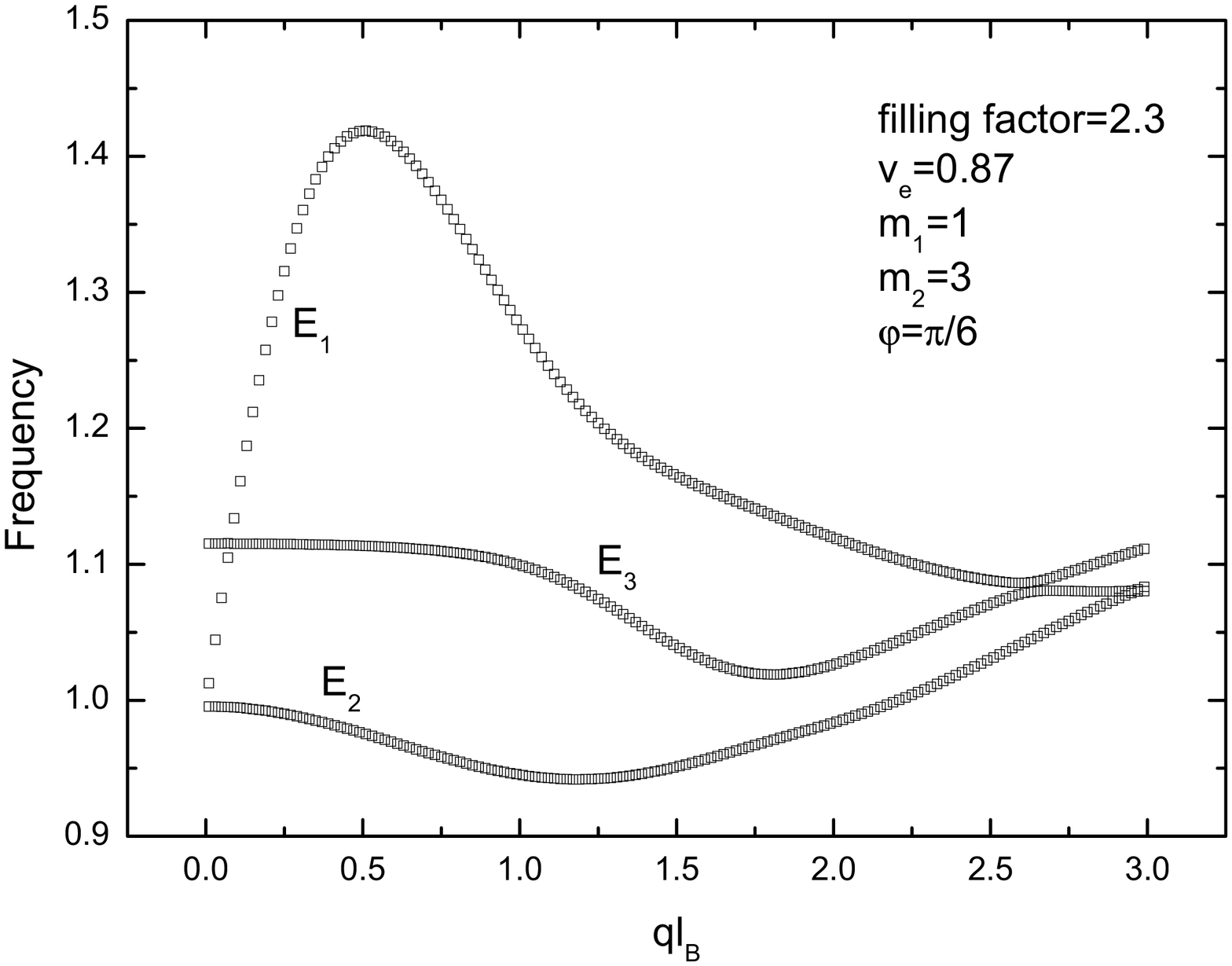}
\caption{ (Color online)
The frequency of magneto-plasmon modes versus the amplitude
of the wave vector, for a non-integer filling factor. }
\end{figure}

\begin{figure}[h]
\includegraphics[angle=90, width=8.truecm]{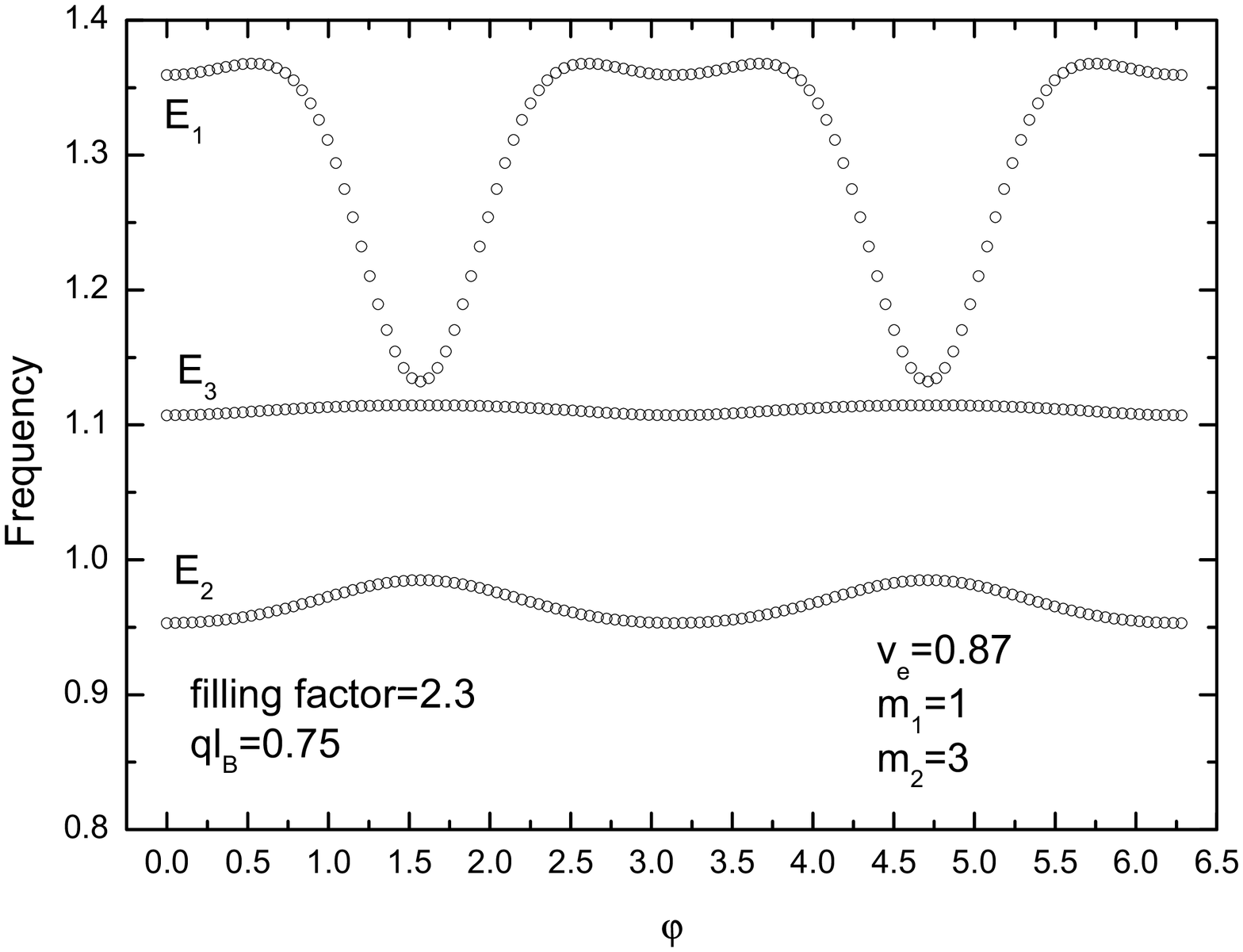}
\caption{ (Color online)
The frequency of magneto-plasmon modes versus the direction
of the wave vector, for a non-integer filling factor. }
\end{figure}

\end{document}